\documentstyle[12pt,aaspp]{article}
\received{}
\accepted{}
\begin{document}

% WHP definitions
%\input epsf
%\epsfxsize=6.75in
%\epsfxsize=5.25in
%\special{psfile="#1" hoffset=-20 voffset=-36 hscale=65 vscale=65}}
\def\pslandinsert#1{\epsffile{#1}}
\def \b0{\Omega_{0}}
\def\eqref#1{\ref{eq#1}}
\def\e#1{\label{eq#1}}
\def\be{\begin{equation}}
\def\ee{\end{equation}}
\def\ast{\mathchar"2203} \mathcode`*="002A
\def\rslash{\backslash} \def\oforder{\sim}
\def\larrow{\leftarrow} \def\rarrow{\rightarrow}
\def\darrow{\Longleftrightarrow}
\def\defeq{\equiv} \def\lteq{\leq} \def\gteq{\geq} \def\neq{\not=}
\def\<={\leq} \def\>={\geq} \def\lsls{\ll} \def\grgr{\gg}
\def\all{\forall} \def\lub{\sqcup} \def\relv{\vert}
\def\leftv{\left|} \def\rightv{\right|}
\def\%{\char'045{}}
\def\_{\vrule height 0.8pt depth 0pt width 1em}
\def\leftbrace{\left\{} \def\rightbrace{\right\}}
\def\sectsign{\S}
\def\prop{\propto}
\newbox\grsign \setbox\grsign=\hbox{$>$} \newdimen\grdimen \grdimen=\ht\grsign
\newbox\simlessbox \newbox\simgreatbox
\setbox\simgreatbox=\hbox{\raise.5ex\hbox{$>$}\llap
     {\lower.5ex\hbox{$\sim$}}}\ht1=\grdimen\dp1=0pt
\setbox\simlessbox=\hbox{\raise.5ex\hbox{$<$}\llap
     {\lower.5ex\hbox{$\sim$}}}\ht2=\grdimen\dp2=0pt
\def\simgreat{\mathrel{\copy\simgreatbox}}
\def\simless{\mathrel{\copy\simlessbox}}
\def\spose{\rlap} \def\caret#1{\widehat #1}
\def\hat#1{\widehat #1} \def\tilde#1{\widetilde #1}
\def\limitswitch{\limits} \def\dispstyle{\displaystyle}
\def\dot#1{\vbox{\baselineskip=-1pt\lineskip=1pt
     \halign{\hfil ##\hfil\cr.\cr $#1$\cr}}}
\def\ddot#1{\vbox{\baselineskip=-1pt\lineskip=1pt
     \halign{\hfil##\hfil\cr..\cr $#1$\cr}}}
\def\dddot#1{\vbox{\baselineskip=-1pt\lineskip=1pt
     \halign{\hfil##\hfil\cr...\cr $#1$\cr}}}
\def\Abf{{\bf A}}\def\ybf{{\bf y}}\def\Ebf{{\bf E}}\def\cbf{{\bf c}}
\def\Zbf{{\bf Z}}\def\Bbf{{\bf B}}\def\Cbf{{\bf C}}\def\bfB{{\bf B}}
\def\bfq{{\bf q}}\def\bfc{{\bf c}}\def\bfy{{\bf y}}\def\bfv{{\bf v}}
\def\bfp{{\bf p}}
% end WHP definitions

\def \GRBs{$\gamma$-ray bursts }
\def \GRBsb{$\gamma$-ray bursts}
\def \GRB{$\gamma$-ray burst }
\def \Paczy{Paczy{\'n}ski }
\def \paczy{Paczy{\'n}ski }
\def \etal{{\it et al. }}
\def \vovmax{$\left<V/V_{max}\right>$}
\def \vovmaxb{$\left<V/V_{max}\right>\,\,$}
\def \cmaxcmin{C_{max}/C_{min}}
\def \CMB{cosmic microwave background}

\title{The Expected Dipole In The Distribution of \\
Cosmological $\gamma$-ray Bursts}

\vspace{0.7in}

\author{Eyal Maoz}
\vspace{0.7in}

\affil{Harvard-Smithsonian Center for Astrophysics, \\
MS 51, 60 Garden Street, Cambridge, MA 02138}

\vspace{0.3in}
\centerline{E-mail: maoz@cfa.harvard.edu}
\vspace{0.8in}
\centerline{$\dagger\,$ Submitted to the {\it Astrophysical journal}}
%\end{document}

\newpage
\begin{abstract}
If \GRBs originate at cosmological distances then their angular
distribution should exhibit a dipole in the direction of the
solar motion relative to the cosmic microwave background. This is due to the
combined
effects of abberation, an anisotropic shift of the burst event rate, and
an angular variation in the distance out to which bursts can be detected.

We derive the amplitude of the expected
dipole for an $\b0\!\le\!1$ cosmological model, and for various possible
evolution rates of the burst population. Although our dimensionless
velocity with respect
to the CMB rest frame is of order $10^{-3}$, the dipole amplitude
is of order $10^{-2}$, an order of magnitude larger.
The results depend very weakly on the value of $\b0$, but are
sensitive to the spectral index of the bursts' photon spectra, and to the
rate of evolution of the burst population.
There is no dependence on the value of the Hubble constant.

A clear detection of the dipole will require a larger sample of \GRBs than
currently available (of order $10^4$ bursts). Future
statistical analyses of the hypothesis that
bursts originate at cosmological distances should take this effect into
account,
 rather than assuming a perfectly isotropic distribution, for obtaining
the correct statistical significance of their results.

\end{abstract}
%%%%%%%%%%%%%%%%%%%%%%%%%%%%%%%%%%%%%%%%%%%%%%%%%%%%%%%%%%%%%%%%%
\newpage

\section{INTRODUCTION}
If the observed \GRBs originate at cosmological distances (e.g., \paczy 1991)
then the distance
scale of their distribution must correspond to a redshift of order unity
for the burst intensity distribution to be consistent
with observations (e.g., \paczy 1991;
Piran 1992; Mao and \paczy 1992; Fenimore \etal 1992;
Wickramasinghe \etal 1993).
Assuming that the universe is homogeneous and free of bulk flows on such
scales the angular distribution of bursts should appear to be perfectly
isotropic to
an observer at rest with respect to the cosmic microwave background (CMB)
frame, up to statistical fluctuations.
However, the solar system is moving relative
to the CMB frame at a speed of $370\!\pm\!10$ km s$^{-1}$ in
the direction $(l,b)\!=\!(264.7\deg,48.2\deg)$ (Peebles 1993).
Consequently, the bursts' distribution should exhibit
a dipole component in that direction due to several effects:
abberation, anisotropic Doppler shift of the event rate, and the
angular variation in the distance out to which bursts can be detected.

We derive the amplitude of the expected dipole for a Friedmann
cosmological model, examine its dependence on various evolution rates
of the burst
population, and discuss its dependence on the luminosity function.
For sets of parameters which are consistent with the observed \vovmaxb
 parameter we obtain a dipole amplitude of $\sim 10^{-2}$, an order of
magnitude larger than that of the CMB temperature dipole.
In \S{2} we derive the three effects which contribute to the
anisotropy, and evaluate the amplitude
of the dipole in \S{3}. In \S{4} we summarize the results and discuss
their implications.

\section{THE ANISOTROPY EFFECTS}
\def \thetab{\tilde{\theta}}
Denoting the dimensionless
velocity of the solar system relative to the CMB frame by
$\beta\!\equiv\!{v/c}$ and using the Lorentz
transformation it can be shown that photon
directions are related by
\be \cos\theta  = {\cos\thetab + \beta \over 1 + \beta\cos\thetab
} \,\,\,\,\,\,\,\,  ,  \ee
where $\thetab$ and $\theta$ are the angles between our direction of motion and
the direction to a source in the CMB rest frame and in our frame, respectively.
Therefore, assuming an isotropic distribution of bursting objects in the CMB
frame, the angular distribution of sources as observed by us
should be proportional to
\be {dN_s\over d\Omega} \, \propto \, f^{2}(\theta)  \:\: , \ee
where
\be f(\theta) \equiv { \sqrt{1-\beta^2} \over 1 - \beta\cos\theta }
\, \simeq 1 + \delta(\theta) \,\,\,\,\, , \ee
$\delta(\theta)\equiv \beta\cos\theta$, and the rightmost approximation
is due to $\beta\!\ll\!1$.
This is the effect of abberation which makes the sources ``bunch up''
in the forward direction.
In addition, the burst frequency is Doppler shifted
by the factor $f(\theta)$, implying a highest event rate
in the direction of motion.
These effects are independent of the cosmological model and of
possible evolution of the burst population.

The part of a burst's intrinsic luminosity which is
shifted into the detector bandwidth depends on the effective Doppler
shift. Thus, the number of detectable events also varies with direction.
 In order to calculate this effect let us assume the following: 1)
a Friedmann cosmological model with vanishing cosmological constant. 2)
all the bursts have identical
power-law spectra for the photon number in the comoving frame,
$n_{\gamma}(E)\!\propto\! E^{-S}$. 3) The burst detector is sensitive to
photons in a fixed energy bandpass, $E_1\!\le\! E\!\le\! E_2$,
and is triggered by a peak flux higher than a given detection treshold
$F_{min}$ (the peak flux is proportional to the peak photon count rate).
We shall also assume that the bursts are ``standard candles'', and discuss
broader luminosity functions in \S{3}.

Let us define the luminosity of a source by
\def \zgal{{\~z}}
\be L \equiv \int_{E_1}^{E_2}{E\, n_{\gamma}(E)\, dE}  \,\,\,\,\,   ,    \ee
where $n_\gamma(E)$ is normalized accordingly. For a detector at rest relative
to the CMB frame the burst's
luminosity which is shifted into the detector bandwidth is
$(1+z)^{2-S}L$, where $z$ is the cosmological
redshift of the source in the CMB frame.
Dividing by $(1+z)^2$ for
the time dilation in the reception of photons and for the loss of
energy per photon, the peak flux at the observer is
\def \zmaxg{{\tilde{z}}_{max}}
\be F(z) = {L\over 4\pi} {(1+z)^{-S} \over r^{2}(z) }
\,\,\,\,\,\,\, , \ee
where $r$ is the proper motion distance to the source and is given by
\be r(z) = {2c\over H_0} \, {\Omega_{0}z +
 (2-\Omega_0)(1-\sqrt{1+\Omega_{0}z}) \over \Omega_{0}^{2} (1+z)}
\:\:\:\:\:. \ee
Thus, in the CMB rest frame bursts with luminosity $L$ can be detected
out to a redshift $\zmaxg$
which is defined by $F(\zmaxg)\!=\!F_{min}$, where $F_{min}$ is the detection
treshold.
In our moving frame, the effective
Doppler shift of photons from a source which is located at a cosmological
redshift $z$ is $(1+z)/f(\theta)$. Therefore, a detector in our frame
can detect bursts out to a cosmological redshift $z_{max}$ which
is defined by
\be { (1+\zmaxg)^{-S} \over r^{2}(\zmaxg) } \,
= \, { (1+z_{max})^{-S}\,\left[f(\theta)\right]^{S}
 \over r^{2}(z_{max}) } \,\,\,\,\,  .  \ee
Obviously, $z_{max}\!=\!z_{max}(\zmaxg,\theta,S)$. Substituting
$z_{max}\!\equiv\! \zmaxg + \Delta z$, replacing
$f^{S}$ by $1\!+\!S\delta(\theta)$ (see Eq. [3]),
and keeping terms up to first order
($\delta\!\ll1$ and consequently $\Delta{z}\!\ll\!\zmaxg$), we obtain
\def \rmax{r_{max}}
\be \Delta z =  \left( {{2\over r(\zmaxg)}{\left
{dr\over dz} \right|_{\zmaxg} \!\!\!\! + \, {S\over 1+\zmaxg} }
\, \right) S\delta(\theta) \,\,\,\,\,\,\,\,\,  ,  \ee
where $r(z)$ is given by equation (6).
Thus, the number of events that we should observe
in a solid angle $d\Omega$ is
\be {dN(\theta) \over d\Omega} \propto
 \left[ 1+\delta(\theta)\right] }^3  \!\!\!\!\!
     \int\limits_{0}^{\zmaxg+\Delta{z(\theta)}} {\!\!\!\!\! {n_s(z)
       \over (1+z)} \,
       r^{2}(z) \, {dr\over dz} \, dz } \,
\,\,\,\,\,\,  , \ee
where $n_s(z)$ is the number of sources per comoving volume at cosmological
redshift $z$, and the factor $[1\!+\!\delta]^3$ is due to the effects of
abberation and the modified event rate that we discussed earlier.

The burst population may evolve since $z\!\sim\!{1}$ until
the present epoch. Let us assume that the comoving
source number density is given by
$n_s\!\propto\!(1+z)^{-\alpha}$, so $\alpha\!=\!0$ corresponds to a
constant rate of bursts per unit
comoving volume per unit comoving time,
and positive values of $\alpha$ describe an increase in the source
population, or equivalently in the intrinsic event rate, with time.
Thus, the integral in equation (9) can be denoted by $T(z_{max},\b0,\alpha)$,
where
\be T(z,\b0,\alpha)\! &\equiv& \!
 \int\limits^{z} {\!
    {r^2 \over (1+z)^{1+\alpha}}\,{dr\over dz}\,dz} \,\,\,\,\,\,\, .  \ee
This integral can be evaluated analytically for certain values of $\alpha$,
e.g., for $\alpha\!\!=\!\!0$ ($n_s\!=\!{\rm constant}$) we
obtain
\begin{eqnarray} T(z,\b0\!<\!1,0) &=&
     {(2-\b0)\left[x-8x^{3}(1-\b0)\right]
\sqrt{1+4x^{2}(1-\b0)}\over 64 (\b0-1)^2}
     - {x^4\over 2}  \nonumber \\ & & \nonumber \\
       & &
   \,\,\,\,\,\,\,\,\,\,\,\,\,\,\,\,\,\,\,\,\,\,\,\,\,\,\,\,\,\,\,\,\,
   \,\,\,\,\,\,\,\,\,\,\,\,\,\,\,\,\,\,
       +\, {\b0 x^3\over 6(\b0-1)} + \,
{(\b0-2){\rm arcsinh}(2x\sqrt{1-\b0}) \over 128 (1-\b0)^{5/2}}  \,\,\,  ;
      \nonumber \\ & & \nonumber \\
T(z,\b0\!=\!1,0) &=& {x^3\over 3} - {x^4\over 2} + {x^5\over 5}
\,\,\,\,\, ,
\end{eqnarray}
where $x\!\equiv\!r(z)H_{0}/2c$, and we have ignored the coefficient
$(2c/H_0)^3$ since equation (9) is a proportionality relation.

We may replace
$T(\zmaxg\!+\!\Delta{z})$ by $T(\zmaxg)+ (dT/dz)|_{\zmaxg}\Delta{z}$ due to
$\Delta{z}\!\!\ll\!\zmaxg$. Thus,
substituting equation (8) for $\Delta{z}$, and using the definition of $\delta
(\theta)$, we obtain
\be {dN(\theta)\over d{\Omega}} \, \propto\,
 1\, +\,  \left(3 + K\right)\! \beta\cos\theta \, + \, O(\beta^2)
  \,\,\,\,\,\,\,  , \ee
where
\be K(\zmaxg,\b0,\alpha,S) \equiv
 \left( {{2S\over r(\zmaxg)}{\left
{dr\over dz} \right|_{\zmaxg} \!\!\!\! + \, {S^{2}\over 1+\zmaxg} }
\, \right)
 {1\over T(\zmaxg)} \left {dT\over dz}\right|_{\zmaxg} \,\,\,\,\,\, ,  \ee
and $T$ is defined by equation (10).
Notice that the above results are
independent of the value of the Hubble constant.
In order to gain some insight  we calculated the function $K$
for the case of $\alpha\!=\!0$ and
found it to be well fitted (to within a few percents) by
\be K \, \simeq \, 6.7\, \left({S\over 2}\right)^{\!1.4} \!
 \b0^{-1/3} \,  \zmaxg^{\,-2.7}
\,\,\,\,\,\,\,\,\,\,\,\,\,\,\,\,\, (\alpha\!=\!0) \ee
in the range of parameters $\,1.0\!\le S\le\!2.5\, ,\: 0.2\!\le\!\b0\!\le1,
\,  $ and $1\!\le\zmaxg\!\le2$.
%
%%%%%%%%%%%%%%%%%%%%%%%%%%%%%%%%%%%%%%%%%%%%%%%%%%%%%%%%%
\section{THE DIPOLE AMPLITUDE}

The redshift out to which bursts are detected, $\zmaxg\,$, is not a free
parameter. It is determined by the requirement that the burst intensity
distribution coincides with the observed one, namely, that the \vovmaxb
 parameter equals the measured value. Since
\vovmax_{BATSE}$=0.330\pm0.016$ (Meegan \etal 1993), the BATSE instrument
can detect bursts out to a redshift $\zmaxg$ which is determined by
\be \langle {V_{   }\over V_{max}}\rangle \equiv
\, {1\over T(\zmaxg,\b0,\alpha)} \int\limits_{0}^{\zmaxg}{
 {F^{3/2}_{min}\over F^{3/2}}
 \, {r^2 \over (1+z)^{1+\alpha}} \, {dr\over dz}\, dz} \, \,= \, 0.330
\,\,\,\,\, , \ee
where $T$ is defined in equation (10),
$F$ is given by equation (5), and $F_{min}\!\equiv\!{F(\zmaxg)}$.
Thus, $\zmaxg$ depends on $\alpha$, $S$, and $\b0$.

There is a considerable diversity in the observed
spectra of \GRBsb, but the average
spectral index of the photon spectrum, $S$, is somewhere between $-1.5$
and $-2$ (schaefer \etal 1992).  Regarding the $\alpha$ parameter,
it is clear that
the population of cosmological \GRBs may evolve with epoch but we have no
observational constraint on that.
Therefore, we shall examine the
cases of moderate ($\alpha\!\!=\!\! 1/2$) and rapid ($\alpha\!\!=\!\!1$) rates
of evolution, as well as the case of
a constant comoving event rate ($\alpha\!=\!0).

Substituting $\beta\!=\!1.233\!\times10^{-3}$ in equation (12),
the amplitude of the dipole is
\be \left[ 3.7 + 1.23K(\zmaxg,\b0,\alpha,S)\right]\times 10^{-3}
\,\,\,\,\, , \ee
where, for a given $\b0$, $\alpha$,
 and $S$, the value of $\zmaxg$ is determined from
equation (15), and $K$ is evaluated using
equation (13). The results for various
combinations of parameters are shown in Table 1.
For reasonable sets of parameters the dipole amplitude is of order $10^{-2}$,
an order of magnitude larger than $\beta$,
and it is almost independent of the value of $\b0$.

At first sight it seems surprising that the dipole amplitude increases when
evolution of the burst population becomes significant (Table 1).
Afterall, a higher
value of $\alpha$ implies a smaller number of bursts originating within a
given range of redshift, $[z,z\!+\!\!\Delta{z}]$. However, an increasing rate
of
evolution also compels a {\it lower\/} $\zmaxg$ since evolution replaces
some of the ``redshift effect'' which is required for a consistency with
the observed
\vovmax. Therefore, the proper volume within $[\zmaxg,\zmaxg\!+\!\Delta
z]$, relative to the volume within $[0,\zmaxg]$ is of order $3\Delta r/r$,
where $r\!=\!r(\zmaxg)$ and $\Delta r\!=\!(dr/dz)|_{\zmaxg}\Delta z$.
Thus, since $\Delta{r}/r\sim O(\beta)$, and the effect of evolution is of order
$\alpha\beta$, the net effect of an
increasing $\alpha$ is an increase in the fraction of detectable bursts,
as long as $\alpha\!\lesssim3$.

We should keep in mind that the possibility of a negative value of
$\alpha$, namely, a decrease in the comoving event rate with time,
cannot be excluded. In such case the dipole amplitude will be lower,
and the redshift out to which bursts are detected
will be higher. We argue that $\alpha$ is unlikely to be negative and
large for the following reasons: 1) bursts at a too high redshift would
introduce a severe difficulty to most progenitor models, e.g., the merging
neutron star model,  since galaxies may
not have formed yet. 2) it would imply a strong correlation between
the brightness and the duration of bursts, which is not observed.

The assumption that all the bursts are ``standard candles'' may be adequate,
but a broader luminosity function cannot be excluded. In such case, $\zmaxg$
would depend on $L$ through equation (5), and an integration over the range
of possible luminosities, $\int{\!dL \,\Phi(L)\,}$, should precede the r.h.s of
equations (9), (10), and (15). We argue that if the luminosity function
is falling with increasing luminosity, e.g., a power law distribution
($\Phi(L)\!\propto\! L^{-\gamma}$), than
the dipole amplitude will {\it increase}, the more so for a larger value of
$\gamma$. The reason for that is the following:
assuming ``standard candles'' implies that most of the observed bursts
originate at distances  close to the boundary of the sphere of detectable
bursts. By contrast, in the case of a steeply falling luminosity function
the average distance to a burst may be considerably smaller.
Therefore,
from an argument similar to the one presented in the previous paragraph,
as well as from the apparently (inverse)
strong dependency of the dipole amplitude on the cosmological redshift
(e.g., equation [12]) we conclude that replacing the ``standard
candle'' assumption by a falling luminosity function will tend to increase the
dipole amplitude. A detailed calculation for specific
luminosity functions is beyond the scope of the present study.

%%%%%%%%%%%%%%%%%%%%%%%%%%%%%%%%%%%%%%%%%%%%%%%%%%%%%%%%%%%%%%%%%%%%
\section{CONCLUSION}

{\it Assuming\/} that \GRBs originate at cosmological distances, we have shown
that three effects combine together to produce a dipole
anisotropy in the bursts' angular distribution. The dipole should point
 in the direction of the solar
motion relative to the cosmic microwave background rest frame.

The amplitude of the predicted
dipole depends weakly on $\b0$, but it is sensitive to the
the spectral index of the photon
spectra, and to the rate of evolution of the burst population. It is
independent of the value of the Hubble constant. The maximum
redshift at which bursts can be detected is not a free parameter but is
constrained by
the requirement that the \vovmaxb parameter be consistent with observations.
The dipole amplitude turns out to be of order $10^{-2}$ for various
combinations of parameters. This is an order of
magnitude larger than what one would expect since the solar
velocity with respect to the CMB is $370\, {\rm km\,
 s}$^{-1}$ ($\simeq10^{-3}$).

Obviously, the sun is in motion relative to the Galaxy too, so one would
expect a similiar effect even if bursts originate within an extended Galactic
halo (e.g., Fishman \etal 1978; Atteia and Hurley 1986; Maoz 1993). However,
in this case
the amplitude of the predicted
anisotropy ($<\!1\%$) is negligible relative to the
uncertainties in our understanding of the exact shape of the halo.
It is only within the cosmological origin hypothesis that the
dipole due to the solar motion is of practical interest.

The predicted dipole cannot provide a strong test to the hypothesis of
a cosmological origin of \GRBs until a sample of the order of $10^{4}$
bursts is established. The sky exposure map will also have to be complete to
a sufficient accuracy. In the near future, being aware of the expected dipole,
rather than testing the consistency of the
data with a perfectly isotropic distribution,
will enable future statistical analyses to derive a more reliable statistical
significance for their results.

I wish to thank Avi Loeb, Ramesh Narayan, and Tsvi Piran for comments.
This work was supported by the U.S. National Science Foundation, grant
PHY-91-06678.

\newpage
\doublespace

\vspace{0.9in}
{\bf TABLE 1.} The Dipole Amplitude
\vspace{0.2in}

\begin{tabular}{l c c| c c|}
source &\multicolumn{2}{c|}{$\b0=1$}
      & \multicolumn{2}{c}{$\b0=0.3$}  \\ \cline{2-3} \cline{4-5}
evolution& S=2 &\multicolumn{1}{c|}{S=1.5} &S=2 &S=1.5 \\ \hline
\!$n_s=$ constant
&$(1.02\, ;\, 11.7\!\times\! 10^{-3})
&$(1.30\, ;\, 6.4\!\times\! 10^{-3})
&$(1.22\, ;\, 11.2\!\times\! 10^{-3})
&$(1.65\, ;\, 6.1\!\times\! 10^{-3})
\\
\!$n_s\propto (1\!+\!z)^{-1/2}$
&$(0.86\, ;\, 14.3\!\times\! 10^{-3})
&$(1.06\, ;\, 7.8\!\times\! 10^{-3})
&$(0.99\, ;\, 13.9\!\times\! 10^{-3})
&$(1.27\, ;\, 7.6\!\times\! 10^{-3})
\\
\!$n_s\propto (1\!+\!z)^{-1}$
&$(0.75\, ;\, 17.4\!\times\! 10^{-3})
&$(0.90\, ;\, 9.4\!\times\! 10^{-3})
&$(0.83\, ;\, 18.0\!\times\! 10^{-3})
&$(1.02\, ;\, 9.6\!\times\! 10^{-3})
\\
\end{tabular}
\vspace{0.3in}

{\bf Table 1} - The redshift out to which bursts can be detected, $\zmaxg$,
 and the amplitude of the dipole component
(equation [16]), evaluated for several combinations of the photon
spectral index, $S$, the cosmological density parameter, $\b0$, and the
rate of evolution of the burst population. In general, $\zmaxg\!\simeq\!1$ and
the dipole amplitude is of order $10^{-2}$. The dependence on the various
parameters is discussed in \S{3}.

\end{document}